# A method to specify X-ray mirrors for nanoprobe beamlines

Authors

**Zengguang Tang[a], Zhenjiang Xing[a]\*, Lei Li[a], Yan Zhang[a], Rongchang Chen[a] and Zheng Tang[a]\***

[a] Institute of Advanced Science Facilities, 268 Zhenyuan Road, Shenzhen, Guangdong, 518107, People's Republic of China

Correspondence email: xingzhenjiang@mail.iasf.ac.cn; tangzheng@mail.iasf.ac.cn

**Synopsis** By using the method that we proposed, one can clearly illustrate the requirements of optical performance for the surface accuracy of reflective mirrors, which avoids procuring significantly more expensive mirrors than necessary. Furthermore, we have studied in depth the problem that power spectral density (PSD) function cannot guarantee an accurate judgement of the mirror quality, we acknowledge that while statistical parameters can uniquely determine the Strehl ratio (SR) value, the SR value as a criterion for judgment is stringent.

**Abstract** The development of synchrotron radiation light sources has been going on for decades. Recently, X-ray free-electron lasers as well as the fourth-generation synchrotron radiation sources based on multiple bend achromat lattices are commissioned and constructed. These advanced light source facilities result in beam with superior performance, such as higher brilliance, higher degree of coherence, smaller achievable focused size. To take full advantage of the unique properties of the source, the superior benefits of synchrotron radiation require high standards for the quality of optics used in beamline construction. In this paper, we took nanoprobe beamline as an example and propose a method to achieve pre-designed specifications of beamline by imposing restrictions on certain parameters of the X-ray reflective mirrors. By using this method, one can clearly illustrate the requirements of optical performance for the surface accuracy of reflective mirrors and determine which specific requirement is more stringent, which avoids procuring significantly more expensive mirrors than necessary, the correspondence between spatial frequencies and statistical parameters would also be understood. Furthermore, we have studied the problem that power spectral density (PSD) function cannot guarantee an accurate judgement of the mirror quality. This method can also be generalized to designs of other beamlines.

## 1. Introduction

In the past decade, an increasing number of fourth-generation synchrotron light sources and X-ray free-electron lasers have been under construction or planning worldwide. At the same time, many third-generation synchrotron light sources that have been operating for several decades have also proposed their own upgrade plans. They aim to achieve more collimated, higher brightness, and higher degree of coherence light sources to meet the demands of next-generation experimental techniques such as X-ray photon correlation spectroscopy, coherent diffraction imaging, propagation-based phase-contrast imaging and ptychography (Pfeiffer, 2017; Shpyrko, 2014).

Due to the characteristics of low refractive index and strong absorption of X-rays in various materials, achieving a higher reflectivity often involves utilizing total reflection of the mirror under grazing incidence geometry. Additionally, x-ray reflective mirror is achromatic in a large range of energy values with near 100% efficiency even at high energies of X-rays. Currently, X-ray reflective mirrors have become the commonly used key optical components in advanced light source beamlines. They have been widely applied in beam collimation, focusing, and deflection. Their optical performance directly affects the flux, size, and quality of the beamline's imaging spot. However, it can be compromised due to imperfections in optical elements, its analysis is always an important part of beamline design.

Generally, from a geometric optics perspective, any surface shape error alters the angle of incidence between the light rays and the mirror, thereby changing their exit path and affecting the size and intensity distribution of the imaging beam spot. The slope error which refers to the lower spatial frequency with relatively longer spatial period than the roughness has become one of the primary technical parameters used to evaluate the surface quality of X-ray reflective mirrors. By considering a simple focused optical model and utilizing the theory of optical transfer matrices, and taking into account the approximation of geometric optics, the broadened spot size (root-mean-square, RMS) resulting from the slope error of the mirror can be approximated as $2q\sigma\_slope$, where q and $\sigma\_slope$ are the image distance and RMS residual slope error, respectively.

For the third-generation (and previous) of synchrotron radiation sources, the electron beam has a relatively larger emittance, resulting in a lower degree of coherence of the emitted X-rays. In most cases, the approximation of geometric optics holds well and the geometrical ray tracing method for the beamline simulation is applicable. When evaluating the surface shape error of X-ray reflective mirrors, by considering only the slope error of the mirror is effective. However, for the fourth-generation synchrotron light sources and X-ray free electron lasers, the emittance of the electron beam is significantly reduced, and the degree of coherence is greatly increased (Sanchez Del Rio *et al.*, 2019). The geometrical optics approximation is no longer applicable, and wave optics theory and wavefront propagation approach must be employed.

Rakitin et al. have recently shown that there are two situations in which the simulation results of geometrical ray tracing and wavefront propagation approach are different (Rakitin *et al.*, 2018). One scenario is the degree of coherence of the source is high, and another is the contribution of slit diffraction. For experimental methods that based on the coherence of X-ray, the corresponding beamlines usually use slits to select fully coherent beams. Evidently, in these cases, the above-mentioned relationship between the broadened spot size and the slope error under the geometric optics approximation is no longer valid. By wave optics simulations or theoretical discussions, many researchers have studied the influence of specific spatial frequencies and have pointed out that for nano-focusing case, lower spatial frequencies have a greater impact and only the slope error is not sufficient to specific the mirror (Yashchuk *et al.*, 2015; Pardini *et al.*, 2015; Shi *et al.*, 2016). However, according to the DABAM database (Sanchez Del Rio *et al.*, 2016), lower spatial frequencies are definitely present and have the highest proportion in real reflective mirrors. And the specific values and boundaries of statistical parameters for a reflective mirror still hasn't been raised, and unable to answer whether the non-statistical parameters are also necessary to uniquely determine the expected optical performance.

In this study, we first briefly outline the surface error model developed by earlier researchers, and then we draw some inferences based on the model and the newly developed wave optics simulation method with nanoprobe beamline as an example. Meanwhile, the requirements for the specifications of the mirrors can be clearly seen. Subsequently, we elaborated using the on-axis Strehl ratio (SR) as a criterion for judgment would exclude some "qualified" mirrors.

**2. The mirror model**

**2.1. The real mirrors in beamline**

Usually, the mirror metrology apparatus such as Long Trace Profiler (LTP) records the angles versus the mirror coordinate x. As these angles are very small, their tangent can be approximated by the angle. Hence heights profile can be obtained by integration the slopes profile. Some statistical parameters are used to characterize the height errors and slope errors. They are the standard deviations of slope error and height error, respectively:

$$\sigma^{\mathrm{h}} = \left[\frac{1}{N}\sum_i (z_i - \bar{z})^2\right]^{\frac{1}{2}}, \sigma^{\mathrm{s}} = \left[\frac{1}{N}\sum_i (\theta_i - \bar{\theta})^2\right]^{\frac{1}{2}}, \qquad (1)$$

Where $\theta_i$, $z_i$ is the slope and height along the mirror length, the bar above the letter represents the mean, which is generally removed before data processing and is therefore zero. The measured slopes profiles include the main curvature of the mirror itself. To obtain significative values, it is necessary to remove the surface profile, which is usually referred to as profile detrending.

The DABAM database provides a library of open-source metrology data of real X-ray mirrors measured at different facilities. This provides access to metrology data, which can be utilized in conjunction with simulation tools to assess the impact of optical surface errors on the performance of an optical instruments.

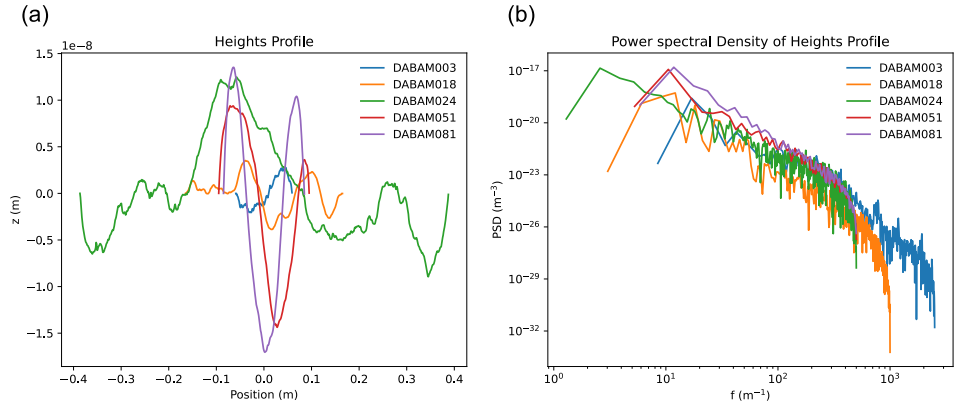

**Figure 1** Several DABAM data. (a) Height error profiles. (b) PSD profiles.

In the database, almost all profile errors of mirror have PSDs that can be approximated as a straight line in the log-log plot, which indicates that PSD decreases with increasing frequency according to a power law. Fig. 1 show several height profiles and their corresponding PSD profiles of DABAM database, and the mirrors are in a variety of surface shape, including spherical, cylindrical, plane, and elliptical. It is noteworthy that the method for calculating the power spectral density function can be found in Reference (Church & Takacs, 1995), which has been incorporated into an American Society for Testing and Materials (ASTM) standard method. In the PSD calculation method, the range of spatial frequency is limited from $f_{min} = 1/ND$ to $f_{max} = 1/2D$, where N is the number of points of the profile and D is the sampling distance. In the PSD plot, there is a step between the first and second frequencies, which is due to the difference in bookkeeping factor where it equals 1/2 at these two extreme frequencies and is equals 1 in other cases.

**2.2. Simulated mirror profile**

As the article reports (Sanchez Del Rio *et al.*, 2016), an ab initio simulated profile with PSD that follows the power law $\propto |f|^{-\beta}$ can be created by a sum of cosine or sinusoidal functions with frequencies in the desired interval and random phase, and with amplitude matching the power law. The accurate formula can be found in Reference (Shi *et al.*, 2016):

$$\Delta h(x) = \sum_n \Delta h_n(x) = \sum_n b_0 n^{s_0/2} \cos\left(\frac{2\pi n}{L} x + \psi_n\right), \qquad (2)$$

where $b_0$ and $s_0$ are two adjustable parameter, the latter determines the slope of the linear fitting of log[PSD(f)] as a function of log(f), n/L is spatial frequency and $\psi_n$ is a random phase shift. The $s_0$ parameter cannot be taken arbitrarily, according to the DABAM database it ranges from -1 to -5.8, which is limited by the manufacturer's process. On the other hand, the value of the $b_0$ parameter is related to the surface error range to be explored. Due to the random phase, the same $b_0$ and $s_0$ do not uniquely determine a surface error profile, but they all have same RMS height error and RMS slope error according to formula (1), the PSD function is also the same.

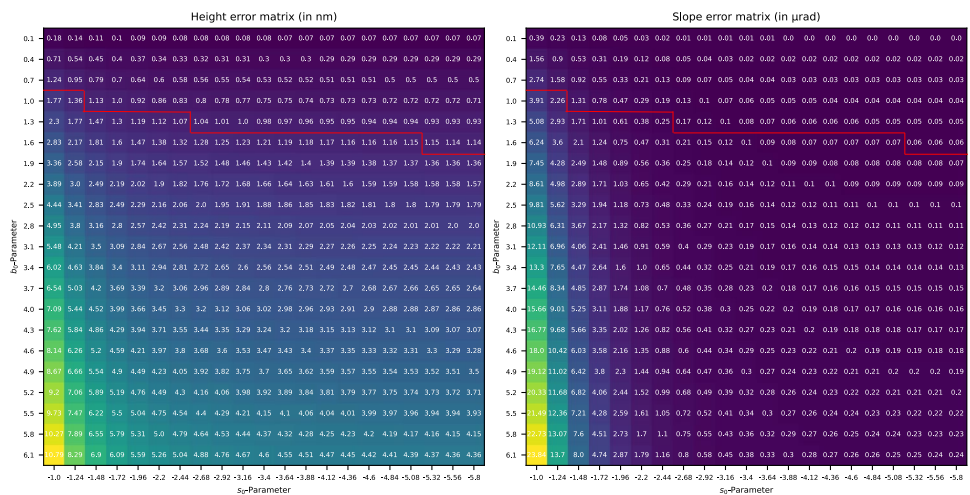

**Figure 2** Simulated surface error results with different $b_0$ and $s_0$ parameters. Left: Height error matrix; Right: Slope error matrix. The red line in the figure will be discussed later.

Fig. 2 shows the simulated surface error results with different $b_0$ and $s_0$ parameters. The values of each row and column in the matrix are monotonically varying and the results of multiple simulations are almost the same, which proves that each pair of parameter represents an infinite number of mirrors, but they have the same RMS slope error, RMS height error. Fig. 3 shows simulated surface error profile by a pair of $b_0$ and $s_0$ parameter.

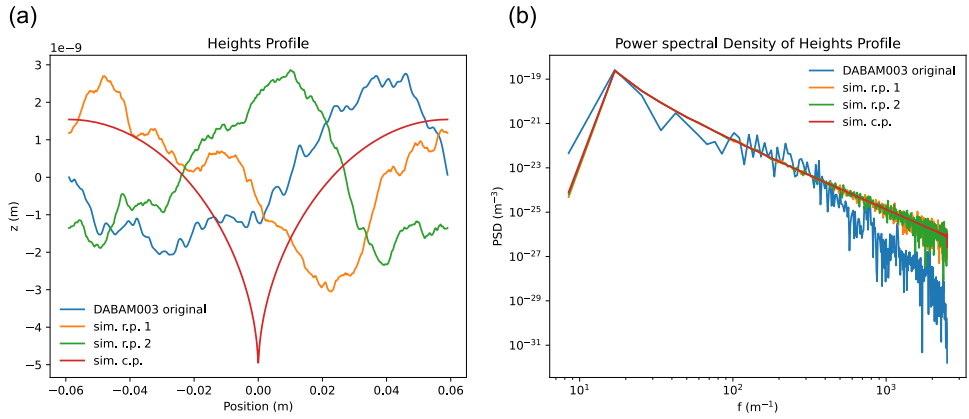

**Figure 3** Simulated surface error profile by a pair of $b_0$ and $s_0$ parameter with the random and constant phase, their $\sigma^h$ and $\sigma^s$ are the same. (a) Height error profiles. (b) PSD profiles.

The surface error profile generated by this model not only have a spatial frequency distribution close to that of a real reflector, but also allows for easy control of variables to study the impact of surface errors. Moreover, it is possible to understand the correspondence between spatial frequencies and statistical parameters, which helps in flexibly specifying to manufacturing vendors. Many researchers cite the work of Church and Takacs that by introducing the concept of critical length, the effect of the face shape error can be divided into two parts, the slope error and the height error can be obtained by integrating the PSD respectively (Pardini *et al.*, 2015; Church & Takacs, 1995; Quan *et al.*, 2008). Shi et al. have pointed out that for the diffraction limited light sources this separation becomes less straightforward (Shi *et al.*, 2016). In addition, different window functions will also affect the results of PSD calculations, so using a table is a reasonably intuitive way to do this.

### 3. Wave optics simulations

The on-axis Strehl ratio (SR) is defined as the ratio between the on-axis intensity of the focal spot of the system under study $I(0)$ and the one produced by the ideal (no surface error, aperture size effect is considered) optical system $I_0(0)$. The closer the ratio is to 1, the better the quality of the image and the smaller the spot extension is.

Fig. 4 is the top view of optical layout of a typical nanoprobe beamline. The optics in nanoprobe beamline are usually designed to achieve a focal spot below 100 nm for resolving the atomic, chemical, and electronic structures of modern advanced materials with very high spatial resolution. To accomplish this, ultra-precise mirrors serve as the nano-focusing mirrors to allow a wavefront preserving transport of photons. Here the used nano-focusing optics are sequential Kirkpatrick-Baes (KB) optics, the effective length (covered by the light beam) of the horizontal focusing mirror (146.326 m from source) is set to 118 mm, its surface error will only affect the horizontal profile of focal spot. The secondary source aperture (SSA) is located at 55 m from the undulator source. Its slit opening is set to provide a fully coherent illumination across the KB optics which can be calculational using a similar equation as described in Reference (Bjorling *et al.*, 2020). The beam intensity profile in the focal plane (146.5 m from source) can be acquired by the wave optics simulations. Up to this point, the SR of any given surface error profile can be predicted.

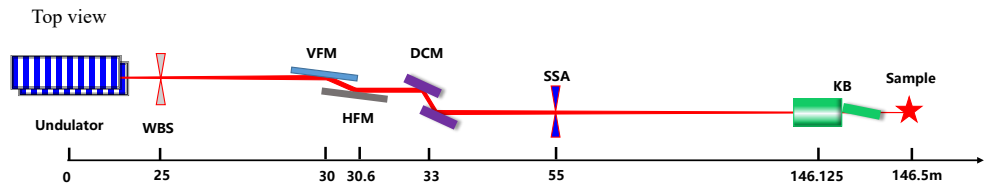

**Figure 4** Scheme of the nanoprobe beamline optical system.

As previously stated, wave optics simulations can provide more buildable results than ray tracing simulations, especially for diffraction-limited synchrotron radiation or X-ray free-electron lasers (FEL) beamlines, which provides a precise method to calculate the SR. For the SRW (Synchrotron Radiation Workshop) code, in order to model and calculation of the light source of the insertion devices, Monte Carlo multi-electron sampling needs to be used and multiple wavefronts must be taken into account to include the effects of partial coherence, which is originated by the electron emittance effects in storage rings (Chubar & Elleaume, 1998; Chubar *et al.*, 2011; Chubar *et al.*, 2017). Together with the wavefront propagation simulation, the whole wave optics simulation of beamline is quite time-consuming in partially coherent light sources. Moreover, the simulation in our case needs to be performed 441 times, which leads to exploration of other approaches. With coherent modes decomposition (CMD) methods, coherence properties of the X-ray beam at any point of the beamline can be completely described by the eigenvalues and coherent modes of the cross spectral density (Glass & Sanchez del Rio, 2017; Xu *et al.*, 2022; Sanchez Del Rio *et al.*, 2022). For these so-called diffraction-limited storage rings, the number of required coherent modes for coherence analysis is about $10^2 - 10^3$, which is much more efficient than the Monte Carlo multi-electron method ($\sim 10^3 - 10^5$). Furthermore, once the undulator is determined, the CMD results of the source would be saved and utilized, avoiding the repeated SR source simulations. Using this method, obtaining the simulated Strehl ratio matrix only took 16 hours on a 56-core computer cluster for simulating 1000 coherence modes.

On the other hand, according to Church's theory, the on-axis SR can be acquired by analytical approach. Critical length needs to be calculated by angular radius of the focal spot. Here we select rectangular window to calculate PSD, the on-axis SR result acquired by analytical approach and wave optics simulations are shown in Fig. 5.

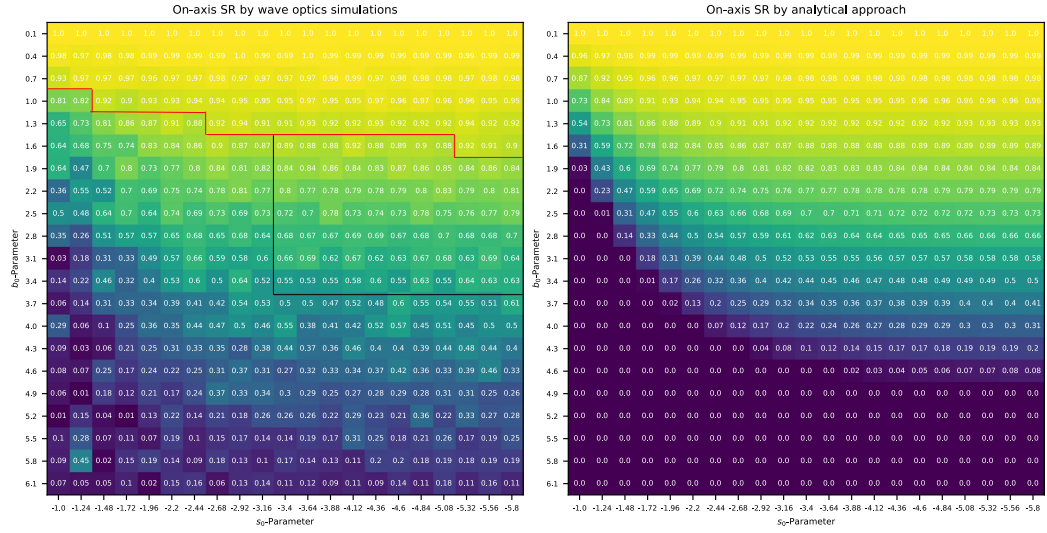

**Figure 5** The on-axis SR result, their coordinates are kept the same as in Fig. 2. Left: Acquired by wave optics simulations; Right: Acquired by analytical approach. The area enclosed by the red line represents SR ≥ 0.9. The black line in the figure will be discussed later.

In On-axis SR matrix by analytical approach, the calculational SR in this case can be negative, but we assign it a value of zero. The same $\sigma^h$ and $\sigma^s$ uniquely determine the value of SR, the SR matrix obtained from the simulation method can also be approximated as such, the values in the lattice are approximately monotonically varying. When the SR values by analytical approach are high (>0.7), they match well with the values obtained from the simulation method. It is also important to mention that the SR values by analytical approach do not depend on the light source of the optical system, in our case the KB mirror receives fully coherent beam, other cases need to verify these conclusions. Based on the SR values obtained from the simulations, we can propose specifications for the mirrors, which need to meet the requirements of beamline. This can be seen as some constraints on the parameters of the mirror. The more grid points are included, the looser the restrictions and the less the cost. Using more than one specification and combining them with logical connectors can include more grid points. As a result, for the horizontal focusing mirror of the nanoprobe beamline example, if SR > 0.9 is selected, the requirement is that height error ≤ 1.07 nm. This single constraint alone contains the maximum number of lattice points and eliminates the need for constraint on slope error and spatial frequency. Even if the slope error exceeds 1 μrad, if the given height error constraints are mainly satisfied, the optical performance is still guaranteed. If only the slope error were used to specify the mirror, the constraint (slope error ≤ 0.05 μrad) would be very strict because it includes relatively few grid points. This agrees with the previous reports, that for the nano-focusing case, the optical performance is not sensitive to the slope error of the nano-focusing mirror. For general optical systems, it can be predicted that both slope and height parameters need to be used to include more lattice points in the SR matrix. Further, one can find the boundary conditions that are both necessary and sufficient for these requirements.

**4. Peak intensity ratio**

For the on-axis SR value, it does not contain all the information about the intensity profiles of the focal spot. When surface error is considered, the highest peak may not be at the zero position. In this case, using the SR value as a criterion for judgment would become stringent, as it does not affect the experiments in the beamline. We can define another reference value, the ratio between the highest intensity of the focal spot of the system under study $I(\max)$ and the one produced by the ideal (no surface error, aperture size effect is considered) optical system $I_0(\max)$, which is called peak intensity ratio (PIR). We are still able to obtain its values by means of wave optics simulations and the simulated results are shown in Fig. 6.

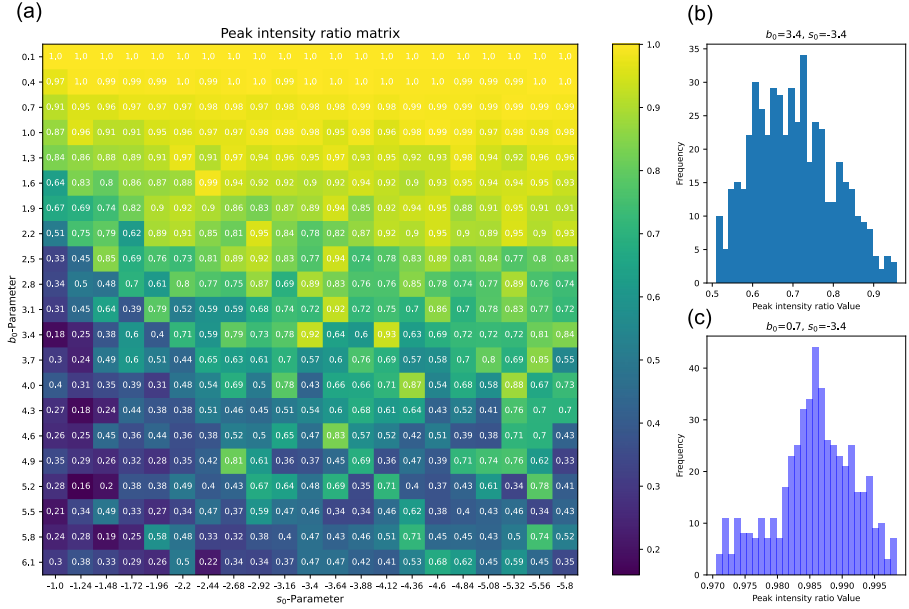

**Figure 6** The PIR results by wave optics simulations. (a) The PIR matrix, their coordinates are kept the same as in Fig. 2. (b) The distribution of PIR value under different grid points. $b_0 = 3.4$, $s_0 = -3.4$; (c) $b_0 = 0.7$, $s_0 = -3.4$.

From Fig. 6(a), we found that the values in the matrix are no longer monotonically changing, with rows having larger $b_0$ parameter being more obvious. This indicates that the same $\sigma^h$ and $\sigma^s$ do not uniquely determine the value of PIR. Then, we take two grid points for 500 simulations respectively, one is $b_0 = 3.4$, $s_0 = -3.4$; the other is $b_0 = 0.7$, $s_0 = -3.4$. The histogram distribution of the data is shown in Fig. 6(b) and Fig. 6(c). The variance of the former is much larger than the latter and it can be inferred that when the data sample is relatively large, the distribution of the PIR value should be Gaussian.

In the case of $b_0 = 3.4$, $s_0 = -3.4$, we selected a few surface error profiles from 500 simulations to study this difference. Fig. 7(a) shows four surface error profiles, all of which have PIR value close to the maximum PIR value in these 500 surface error profiles and Fig. 7(b) is the opposite. According to the previous results, these eight surface error profiles have the same $\sigma^h$, $\sigma^s$ and PSD profiles, but their PIR differ greatly. And these surface error profiles which have PIR value close to the maximum or close to the minimum have their own common characteristics.

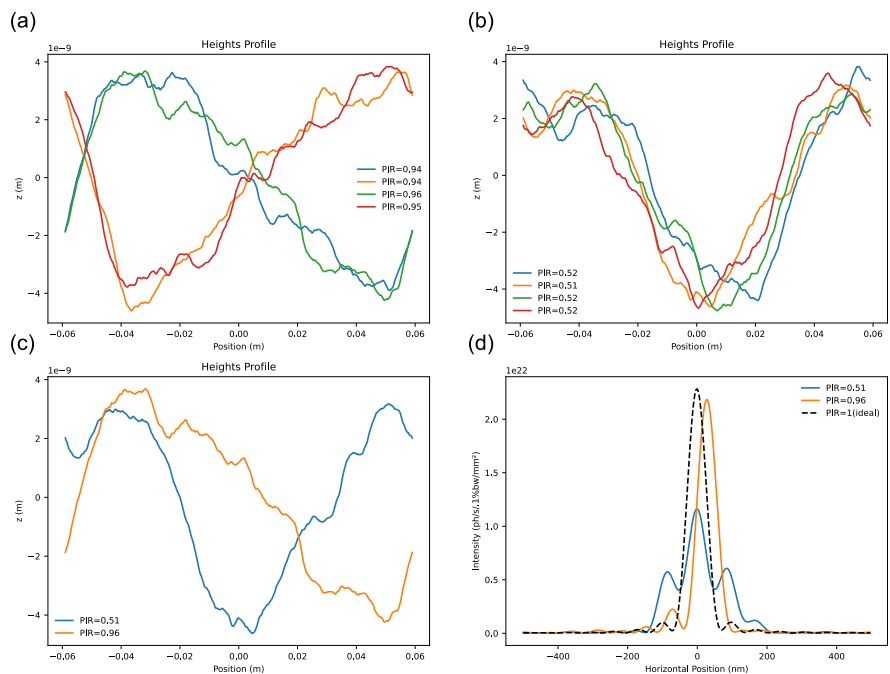

**Figure 7** The simulated results in the case of $b_0 = 3.4$ and $s_0 = -3.4$. (a) Four surface error profiles in with PIR close to maximum. (b) Four surface error profiles with PIR close to minimum. (d) The beam intensity profiles simulated with surface errors profile in (c).

Fig. 7(d) shows simulated spot intensity profiles at sample (focus) by choosing the corresponding surface error profiles of Fig. 7(c). It can be clearly seen that despite the significant difference in their PIR values, their SR values are almost the same. This indicates using SR as a criterion is relatively strict and it is not a necessary condition to ensure the performance of beamline optical. But on the other hand, the PIR value cannot be uniquely determined by the statistical parameters which originate from the fact that their phase distributions are randomized. Hu et al. also observed this phenomenon that even with the same PSD, the focal shape can be quite different, and pointed out that this is due to the different behaviors of sinusoidal and cosinusoidal terms according to equations that generate surface error profiles, in fact, it is related to phase (Hu *et al.*, 2020). Actually, this "quite different" also has some interesting phenomena, as discussed below.

As Fig. 8 shows that, for the surface error profile with equal to 0.51 in Fig. 7(c), compared to the theoretical focal position, the focus is shifted towards the mirror. When examining the actual focal position, the intensity distribution is different from the theoretical focal position, with only a weak side peak, and PIR value is greater than 0.95. The horizontal and vertical focusing mirrors of the KB mirror each affect the intensity distribution in their respective directions. Therefore, in practical engineering, by adjusting the front and rear positions of the horizontal focusing mirror, it is still possible to make the two focuses coincide. In this case, a mirror with a lower PIR or SR is still likely to meet the requirements of optical performance in beamlines. The area enclosed by the black line in Fig. 5 could potentially exhibit this behavior, as they possess a smaller $\sigma^h$ and $\sigma^s$. However not all surface error profiles with lower SR are able to achieve high PIR by moving the focus position. Further simulations indicate that the actual focal point may also shift towards a longer focal length direction. As Fig. 8(c) and (d) shows, in the same scenario, $b_0 = 3.4$, $s_0 = -3.4$, there is a surface error with PIR equal to 0.59. Even when the focal position is adjusted, its side peak remains quite strong. According to this perspective, the mirror with this error profile would be considered substandard.

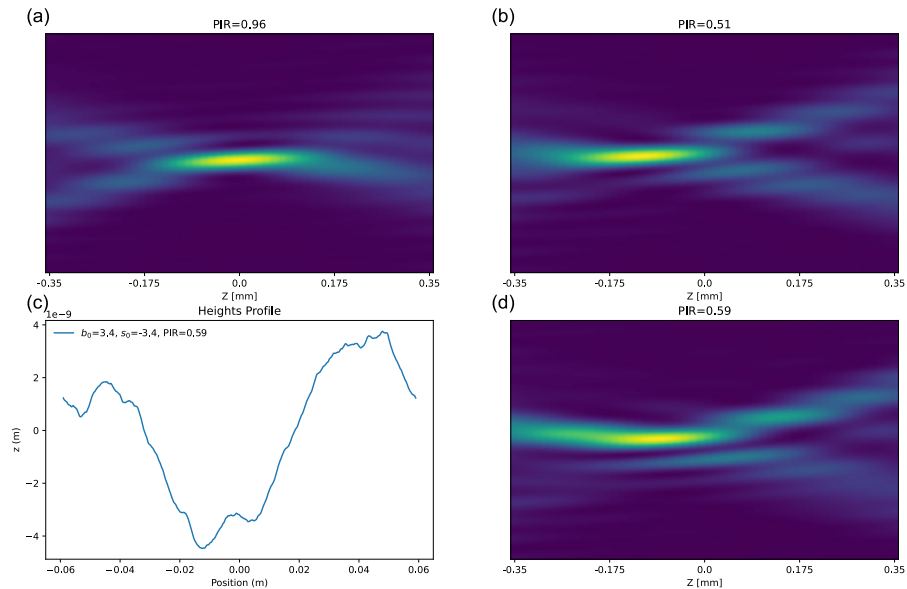

**Figure 8** (a) and (b) Simulations results for caustic with different surface error profiles in Fig. 7(c). (d) Simulations results for caustic with surface error profile in (c).

Now there is a question, how to filter out the surface error profiles that are "qualified" although their SR is low. In fact, the answer has already been provided, that is, these surface error profiles need to be similar to the eight error profiles shown in Fig. 7(a) and (b), their PIR values should be close to the maximum or minimum. We know that the difference of the 8 surface error profiles stems from the random phases in the model, and although we provided a uniform distribution, due to the small sample, their distributions are still different. Further study shows that the surface error profiles in Fig. 7(a) has the fewest phase values approaching zero, while it is the opposite in Fig. 7(b). However, the demand for manufacturers to meet specific phase requirements may not be reasonable. In summary, for the area surrounded by the red line in Fig. 5, the optical performance can be ensured by specifying only statistical parameters of surface error, but the area surrounded by the black line requires additional phase constraints.

## 5. Conclusion and discussion

For an existing reflective mirror, it is relatively easy to evaluate its impact on optical performance. However, before purchasing or manufacturing, we need to provide the corresponding parameters to the manufacturers. In this paper, we proposed an intuitive method, ones can systematically study the effects of different variables and clearly know the requirements for the mirrors, either statistical parameters or spatial frequency constraints. Additionally, we acknowledge that while statistical parameters can uniquely determine the SR value, the SR value as a criterion for judgment is stringent. The effect of surface error on depth of focus has also been systematically studied. This provides another way for manufacturers to produce qualified mirrors and assist in the development of emerging adaptive X-ray mirrors (Goto *et al.*, 2018; Shi *et al.*, 2022). The approach is not very time-consuming, with the emerging wavefront optical simulation methods, ordinary workstations can complete the task within a day.

**Acknowledgements**    This work was supported by Institute of Advanced Science Facilities (IASF), a major national science and technology infrastructure in China.